\documentclass[12pt,onecolumn]{IEEEtran}
\usepackage{amsmath,amssymb,amsfonts,amsthm}
\usepackage{algorithmic}
\usepackage{graphicx}
\usepackage{moreverb}

\usepackage{cite}
\usepackage[hidelinks]{hyperref}
\usepackage{subfigure}

\newtheorem{assumption}{Assumption}
\newtheorem{remark}{Remark}
\newtheorem{definition}{Definition}
\newtheorem{theorem}{Theorem}
\newtheorem{lemma}{Lemma}
\newtheorem{corollary}{Corollary}

\newcommand{\RR}{\mathbb{R}}

\newcommand{\ones}{\mathbf{1}}
\newcommand{\zeros}{\mathbf{0}}
\newcommand{\bA}{\mathbf{A}}
\newcommand{\bB}{\mathbf{B}}

\newcommand{\bP}{\mathbf{P}}
\newcommand{\bS}{\mathbf{S}}
\newcommand{\bT}{\mathbf{T}}

\newcommand{\cA}{\mathcal{A}}
\newcommand{\cC}{\mathcal{C}}
\newcommand{\cE}{\mathcal{E}}

\newcommand{\cG}{\mathcal{G}}
\newcommand{\cW}{\mathcal{W}}
\newcommand{\cV}{\mathcal{V}}
\newcommand{\cL}{\mathcal{L}}

\newcommand{\cI}{\mathcal{I}}

\begin{document}


\title{Cluster synchronization of coupled systems with nonidentical linear dynamics$^\ddag$}

\author{Zhongchang Liu{$^{1,*,\dag}$}, Wing Shing Wong$^{2}$ and Hui Cheng$^{1}$
\thanks{$^1$School of Data and Computer Science, Sun Yat-sen University, Guangzhou 510006, P.R. China}
\thanks{$^{2}$Department of Information Engineering, The Chinese University of Hong Kong, Shatin, N.T., Hong Kong}%
\thanks{$^*$Correspondence to: Zhongchang Liu, School of Data and Computer Science, Sun Yat-sen University, Guangzhou 510006, P.R. China.}%
\thanks{$^\dag$E-mail: zcliu@foxmail.com}%
\thanks{$^\ddag$This is the accepted version of the following article: [Liu Z, Wong W S, Cheng H. Cluster synchronization of coupled systems with nonidentical linear dynamics. International Journal of Robust and Nonlinear Control, 2017, 27(9): 1462--1479], which has been published in final form at [https://onlinelibrary.wiley.com/doi/10.1002/rnc.3811 ]. This article may be used for non-commercial purposes in accordance with Wiley Terms and Conditions for Use of Self-Archived Versions. This article may not be enhanced, enriched or otherwise transformed into a derivative work, without express permission from Wiley or by statutory rights under applicable legislation.
Copyright notices must not be removed, obscured or modified. The article must be linked to Wiley's version of record on Wiley Online Library and any embedding, framing or otherwise making available the article or pages thereof by third parties from platforms, services and websites other than Wiley Online Library must be prohibited.%
}
}
\maketitle

\begin{abstract}
This paper considers the cluster synchronization problem of generic linear dynamical systems whose system models are distinct in different clusters. These nonidentical linear models render control design and coupling conditions highly correlated if static couplings are used for all individual systems. In this paper, a dynamic coupling structure, which incorporates a global weighting factor and a vanishing auxiliary control variable, is proposed for each agent and is shown to be a feasible solution. Lower bounds on the global and local weighting factors are derived under the condition that every interaction subgraph associated with each cluster admits a directed spanning tree. The spanning tree requirement is further shown to be a necessary condition when the clusters connect acyclically with each other. Simulations for two applications, cluster heading alignment of nonidentical ships and  cluster phase synchronization of nonidentical harmonic oscillators, illustrate essential parts of the derived theoretical results. Copyright \copyright 2017 John Wiley \& Sons, Ltd.
\end{abstract}


\vspace{-6pt}

\section{Introduction}
\vspace{-2pt}
Understanding the interaction of coupled individual systems continues to receive interest in the engineering research community \cite{Cao2013}.
The problem of complete synchronization or consensus has been studied for more than a decade, e.g., \cite{Ren2005,Ren2008,Scardovi2009,Li&GRChen2010,Ma&JFZhang2010} to name a few. And application areas include synchronization of coupled harmonic oscillators \cite{Ren2008,Scardovi2009}, formation flying of spacecrafts\cite{Li&GRChen2010}, time synchronization in wireless sensor networks \cite{He2014Time}, and energy management in a smart grid \cite{Zhao&He2016}.
Recently, more attention has been drawn to cluster synchronization problems that study multi-group local interactions. This problem requires individual systems belonging to the same cluster to achieve synchronization while different clusters can achieve distinct synchronized states. Since each system can be affected by systems belonging to external clusters, how to achieve synchronization in each group is a nontrivial extension of consensus problems. The cluster synchronization problem also has wide applications, such as segregation of a robotic team \cite{Kumar2010} or physical particles \cite{Chen2012} into small subgroups, predicting opinion dynamics in social networks \cite{Aeyels2009nonidentical}, and cluster phase synchronization of coupled oscillators \cite{Qin2004CS,Belykh2008cluster}.

In the models reported in most of the literature, the clustering pattern is predefined and fixed; research focuses are on deriving conditions that can enforce cluster synchronization for various system models \cite{Yu_Wang2009group,Yu_Wang2010,Xia2011,Feng2014group,Han&Chen2013,Han&Chen2015,Wu&Chen2009,Sun&Bai2011,Lu&Chen2010,Qin&Yu2013,Yu&Qin2014}. Preliminary studies in  \cite{Yu_Wang2009group,Yu_Wang2010,Xia2011,Feng2014group} reported algebraic conditions on the interaction graph for coupled agents with simple integrator dynamics. Subsequently, a cluster-spanning tree condition is used to achieve intra-cluster synchronization for first-order integrators (discrete time \cite{Han&Chen2013} or continuous time \cite{Han&Chen2015}), while inter-cluster separations are realized by using nonidentical feed-forward input terms. 
For more complicated system models, e.g., nonlinear systems (\cite{Wu&Chen2009,Sun&Bai2011,Lu&Chen2010}) and generic linear systems (\cite{Qin&Yu2013,Yu&Qin2014}), both control designs and inter-agent coupling conditions are responsible for the occurrence of cluster synchronization. For coupled nonlinear systems, e.g., chaotic oscillators, algebraic and graph topological clustering conditions are derived for either identical models (\cite{Wu&Chen2009}) or nonidentical models (\cite{Sun&Bai2011,Lu&Chen2010}) under the key assumption that the input matrix of all systems is identical and it can stabilize the system dynamics of all individual agents via linear state feedback (i.e., the so-called QUAD condition). For identical generic linear systems which are partial-state coupled \cite{Qin&Yu2013,Yu&Qin2014}, a stabilizing control gain matrix solved from a Ricatti inequality is utilized by all agents, and agents are pinned with some additional agents so that the interaction subgraph of each cluster contains a directed spanning tree. 

The system models introduced above can describe a rich class of applications for multi-agent systems. A common characteristic is that the uncoupled system dynamics of all the agents can be stabilized by linear state feedback attenuated by a unique matrix (i.e., static state feedback)  \cite{Qin&Yu2013,Yu&Qin2014}. This simplification allows the derivation of coupling conditions to be independent of the control design of any agent, and thus offers scalability to a static coupling strategy. This kind of benefit still exists for nonidentical nonlinear systems (\cite{Sun&Bai2011,Lu&Chen2010}) which are full-state coupled, since all the system dynamics can be constrained by a common Lipchitz constant (Lipchitz can imply the QUAD condition \cite{DeLellis2011}). However, for the class of partial-state coupled nonidentical linear systems, the stabilizing matrices for distinct linear system models are usually different. It follows that if conventional static couplings (e.g. those in \cite{Qin&Yu2013,Yu&Qin2014}) are utilized, the required conditions for the interaction graph will be correlated with the control designs of all individual systems, and even worse these conditions may never be satisfied for some system models (These points will become clear in Remark \ref{rem:comparisons to identical} of the main part). Therefore, new coupling strategies should be designed so as to cope with the nonidentical system parameters. 

The goal of this paper is to achieve state cluster synchronization for partial-state coupled nonidentical linear systems, where agents with the same uncoupled dynamics are supposed to synchronize together. This is a problem of practical interest, for instance, maintaining different formation clusters for different types of interconnected vehicles, providing different synchronization frequencies for different groups of clocks using coupled nonidentical harmonic oscillators, reaching different consensus values for people with different opinion dynamics, and so on. 
In order to tackle the issues raised by using the conventional \emph{static} couplings,  this paper proposes to use couplings with a \emph{dynamic} structure that incorporates a vanishing auxiliary variable to facilitate interactions among connected agents. With the proposed dynamic couplings, an algebraic necessary and sufficient condition is derived to check the cluster synchronizability of a nonidentical linear multi-agent system. This newly derived algebraic condition is independent of the control design of any agent, and can subsume those published results for integrator systems in \cite{Yu_Wang2009group,Yu_Wang2010,Xia2011,Feng2014group} as special cases. Due to the entanglement between nonidentical system matrices and the parameters from the interaction graph, the algebraic condition may not be straightforward to check. Thus, a graph topological interpretation of the algebraic condition is provided which requires that the interaction subgraph associated with each cluster contains a directed spanning tree. This spanning tree condition is further shown to be a necessary condition when the clusters and the inter-cluster links form an acyclic structure. This conclusion reveals the indispensability of direct links among agents belonging to the same cluster under such special inter-cluster structures, and further strengthens the sufficiency statement presented initially in \cite{Qin&Yu2013}. We also derived lower bounds for the local coupling strengths in different clusters, which are independent of the control designs of any agent thanks to the dynamic coupling structure. Using the commonly used static coupling structures as in \cite{Qin&Yu2013,Yu&Qin2014}, these lower bounds may need centralized computation and may even have no feasible solutions at all. 
The derived results in this paper are illustrated by simulation examples for two applications: cluster heading alignment of nonidentical ships and cluster phase synchronization of nonidentical harmonic oscillators. 

The organization of this paper is as follows: Following this section, the problem formulation is presented in Section \ref{sec:Problemstatement}. In Section \ref{sec:syn_leaderless}, both algebraic and graph topological conditions for cluster synchronization
are developed. Applications of this work and simulation examples are provided in Section \ref{sec:simulation}. Concluding remarks and discussions for future investigations follow in Section \ref{sec:conclusion}.

\section{Problem Statement}\label{sec:Problemstatement}
Consider a multi-agent system consisting of $L$ agents, indexed by $\cI=\{1,\ldots,L\}$, and $N\leq L$ clusters.
Let $\cC=\{\cC_1,\ldots,\cC_N\}$ be a nontrivial partition of $\cI$, that is, $\bigcup_{i=1}^{N}\cC_i=\cI$, $\cC_i\neq\emptyset$, and $\cC_i\cap\cC_{j}=\emptyset$, $\forall i\neq j$.
We call each $\cC_i$ a cluster. Two agents, $l$ and $k$ in $\cI$, belong to the same cluster $\cC_i$ if $l\in\cC_i$ and $k\in\cC_i$.
Agents in the same cluster are described by the same linear dynamic equation:
\begin{equation}
\label{sys:linearmodel}
\dot{x}_l(t) = A_ix_l(t)+B_iu_l(t),\; l\in \cC_i,\; i=1,\ldots,N
\end{equation}
where $x_l(t)\in \RR^{n}$ with initial value, $x_l(0)$, is the state of agent $l$ and $u_l(t)\in \RR^{m_i}$ is the control input;  $A_i\in\RR^{n\times n}$ and $B_i\in\RR^{n\times m_i}$ are constant system matrices which are distinct for different clusters.
\subsection{Interaction graph topology and graph partitions}
A directed interaction graph $\cG=(\cV,\cE,\cA)$ is associated with system (\ref{sys:linearmodel}) such that each agent $l$ is regarded as a node $v_l\in\cV$, and a link from agent $k$ to agent $l$ corresponds to a directed edge $(v_k,\; v_l)\in\cE$. An agent $k$ is said to be a neighbor of $l$ if and only if $(v_k,\; v_l)\in\cE$. 
The adjacency matrix $ \cA=[a_{lk}]\in\RR^{L\times L}$ has entries defined by: $a_{lk}\neq 0$ if $(v_k,\; v_l)\in\cE$, and $a_{lk}=0$ otherwise. In addition, let $a_{ll}=0$ to avoid self-links. Note that $a_{lk}<0$ means that the influence from agent $k$ to agent $l$ is \emph{repulsive}, while links with $a_{lk}>0$ are \emph{cooperative}.
Define $\cL=[b_{lk}]\in\RR^{L\times L}$ as the Laplacian of $\cG$, where $b_{ll}=\sum_{k=1}^{L}a_{lk}$ and $b_{lk}=-a_{lk}$ for any $k\neq l$.

\begin{figure}[ht]
  \centering
  \includegraphics[width=3.8cm]{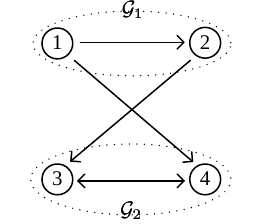}
  \caption{A graph topology partitioned into two subgraphs.}
  \label{fig:topology_illu_leaderless}
\end{figure}

Corresponding to the partition $\cC=\{\cC_1,\ldots,\cC_N\}$, a subgraph
$\cG_i$, $i=1,\ldots,N$, of $\cG$ contains all the nodes with indexes in $\cC_i$, and the edges connecting these nodes. See Figure \ref{fig:topology_illu_leaderless} for an illustration.
Without loss of generality, we assume that each cluster $\cC_i$, $i=1,\ldots, N$, consists of  $l_i\geq 1$ agents ($\sum_{i=1}^{N}l_i=L$), such that $\cC_1=\{1,\ldots, l_1\}$, $\ldots$, $\cC_i=\{\sigma_i+1,\ldots, \sigma_i+l_i\}$, $\ldots$, $\cC_N=\{\sigma_N+1,\ldots, \sigma_N+l_N\}$ where $\sigma_1=0$ and $\sigma_i=\sum_{j=1}^{i-1}l_j, \;2\leq i\leq N$. Then, the Laplacian $\cL$ of the graph $\cG$ can be partitioned into the following form:
\begin{equation}
\label{var:laplacianpartition}
\cL=\begin{bmatrix}
L_{11}&L_{12}&\cdots&L_{1N}\\
L_{21}&L_{22}&\cdots&L_{2N}\\
\vdots&\vdots&\ddots&\vdots& \\
L_{N1}&L_{N2}&\cdots&L_{NN}\\
\end{bmatrix},
\end{equation}
where each $L_{ii}\in \RR^{l_i\times l_i}$ specifies intra-cluster couplings and each $L_{ij}\in \RR^{l_i\times l_j}$ with $i\neq j$, specifies inter-cluster influences from cluster $\cC_j$ to $\cC_i$, $i,j=1,\cdots,N$. Note that $L_{ii}$ is not the Laplacian of $\cG_i$ in general.

This paper will show that both inter-cluster and intra-cluster couplings are important in resulting cluster synchronization. To describe inter-cluster structures, we construct a new graph by collapsing every subgraph $\cG_i$ of $\cG$ into a single node,
and define a directed edge from node $i$ to node $j$ if and only if there exists a directed edge in $\cG$ from a node in $\cG_i$ to a node in $\cG_j$. We say $\cG$ admits an acyclic partition with respect to $\cC$, if the newly constructed graph does not contain any cyclic components.
If the latter holds, by relabeling the clusters and the nodes in $\cG$, we can represent the Laplacian $\cL$ in a lower triangular form
\begin{equation}
\label{var:laplacianpartition_triang}
\cL=\begin{bmatrix}
L_{11}&      &\zeros\\
\vdots&\ddots&      \\
L_{N1}&\cdots&L_{NN}\\
\end{bmatrix},
\end{equation}
so that each cluster $\cC_i$ receives no input from clusters $\cC_j$ if $j>i$. In Figure \ref{fig:topology_illu_leaderless}, the two subgraphs $\cG_1$ and $\cG_2$ illustrate an acyclic partition of the whole graph.

\subsection{The cluster synchronization problem}
The main task in this paper is to achieve cluster synchronization for the states of systems in \eqref{sys:linearmodel} via distributed couplings through the control inputs $u_l(t)$. These controls have dynamic structures as defined below: for $l\in\cC_i$, $i=1,\ldots,N$
\begin{subequations}\label{sys:controllaws_leaderless}
\begin{align}
\label{sys:controllaws_a}
u_l(t)&=K_i \eta_l(t)\\
\label{sys:controllaws_b}
\dot{\eta}_l(t) &= (A_i+B_iK_i)\eta_l+c \left[c_i\sum_{k\in\cC_i}a_{lk}(\eta_k-\eta_l + x_l- x_k)+\sum_{k\notin\cC_i}a_{lk}(\eta_k-\eta_l + x_l- x_k)\right],
\end{align}
\end{subequations}
where $K_i$ is the control gain matrix to be specified; the vector $\eta_l(t)\in\RR^n$, $l\in\cI$ is an auxiliary control variable with initial value, $\eta_l(0)$; $c>0$ is the global weighting factor for the whole interaction graph $\cG$; each $c_i>0$ is a local weighting factor used to adjust the intra-cluster coupling strength of cluster $\cC_i$, $i=1,\ldots,N$.

\begin{remark}
The above control input of each agent uses linear couplings similarly to those static couplings (e.g., those in \cite{Qin&Yu2013,Yu&Qin2014,Wu&Chen2009,Sun&Bai2011,Lu&Chen2010}) which don't use the variables $(\eta_k-\eta_l)$. These linear couplings are distributed and are easy to implement. The introduction of the auxiliary variables $\eta_l(t)$ to form a dynamic structure is partially motivated by the dynamic controllers used for achieving complete synchronization of identical linear systems in \cite{Scardovi2009}. This strategy can convert the problem of synchronizing $x_l$'s to the problem of synchronizing $\eta_l-x_l$, and thus will provide more degrees of freedom to cope with nonidentical system parameters as will become clear in the main part of this paper. The reasons why conventional static couplings (e.g., those in \cite{Qin&Yu2013,Yu&Qin2014,Wu&Chen2009,Sun&Bai2011,Lu&Chen2010}) are not utilized for nonidentical linear systems will also be explained in details in the main part (see Remark \ref{rem:comparisons to identical}). The global weighting factor $c$ is expected to provide large enough coupling strength against the individual system models. The local weighting factors $c_i$'s are supposed to provide large enough intra-cluster coupling strengths against inter-cluster influences. So, the lower bounds of these weighting factors will be presented along with the main results derived in the sequel.
\end{remark}

The cluster synchronization problem is defined below.
\begin{definition}[\cite{Xia2011}]\label{def:cluster_syn}
A linear multi-agent system in \eqref{sys:linearmodel} with couplings in \eqref{sys:controllaws_leaderless} is said to achieve $N$-cluster synchronization with respect to the partition $\cC$ if the following holds: for any $x_l(0)$ and $\eta_l(0)$, $l\in\cI$,
$\lim_{t\rightarrow\infty}\|x_{l}(t)-x_{k}(t)\|=0$ $\forall k,l \in\cC_i$, $i=1,\ldots,N$, $\lim_{t\rightarrow\infty} \eta_l(t)= 0$ $\forall l\in\cI$, and for any set of $x_l(0)$, $l\in\cI$ there exists a set of $\eta_l(0)$, $l\in\cI$ such that $\lim \sup_{t\rightarrow\infty} \|x_{l}(t)-x_{k}(t)\|> 0$ $\forall l\in\cC_i$, $\forall k\in\cC_j$, $\forall i\neq j$.
\end{definition}

By this definition, the system states of agents in the same cluster will synchronize together (i.e., achieve intra-cluster synchronization) while the system states of agents in distinct clusters will be different (i.e., realize inter-cluster separations). In comparison with the definitions in existing papers (e.g., \cite{Xia2011}), the above definition has the extra requirement that all auxiliary variables, $\eta_{l}(t)$, $l\in\cI$ decay to zero so as to guarantee that the control effort of every agent is essentially of finite duration. Further note that intra-cluster state synchronization is required for all $x_l(0)\in\RR^n$ and $\eta_l(0)\in\RR^n$, $l\in\cI$, but inter-cluster separation is only required for all $x_l(0)\in\RR^n$. This is because state separations cannot be guaranteed for any set of $x_l(0)$'s and $\eta_l(0)$'s; an obvious example is that all system states will stay at zero when $x_l(0)=\eta_l(0)=0$ for all $l\in\cI$.

Some assumptions throughout the paper are in order.
\begin{assumption}\label{assump:stabilizable}
Each of the pairs $(A_i,\;B_i)$, $i=1,\ldots,N$ is stabilizable.
\end{assumption}
\begin{assumption}\label{assump:A_unstable}
Each $A_i$ has at least one eigenvalue on the closed right half plane.
\end{assumption}

Note that Assumption \ref{assump:stabilizable} is a necessary condition for achieving consensus for linearly coupled unstable linear multi-agent systems \cite{Ma&JFZhang2010}. Assumption \ref{assump:A_unstable} excludes trivial scenarios where all system states synchronize to zero.
To deal with stable $A_i$'s, one may introduce distinct feed-forward terms in $u_l(t)$ as studied in \cite{Han&Chen2013,Han&Chen2015}.
In order to segregate the system states according to the uncoupled system dynamics in \eqref{sys:linearmodel}, the system matrices $A_i$'s are assumed to satisfy an additional mild condition, namely, they can produce distinct trajectories; rigorously speaking, for any $i\neq j$, the solutions $x_i(t)$ and $x_j(t)$ to the linear differential equations $\dot x_i(t)=A_i x_i(t)$ and $\dot x_j(t)=A_j x_j(t)$, respectively satisfy $\lim \sup_{t\rightarrow\infty} \|x_{i}(t)-x_{j}(t)\|> 0$ for almost all initial states $x_i(0)\in\RR^n$ and $x_j(0)\in\RR^n$.
\begin{assumption} \label{assump:zero_row_sums}
Every block $L_{ij}$ of $\cL$ defined in \eqref{var:laplacianpartition} has zero row sums, i.e., $L_{ij}\ones_{l_j}=\zeros$.
\end{assumption}

This assumption guarantees the invariance of the clustering manifold $\{x(t)=[x_1^T(t),\ldots,x_L^T(t)]^T: x_1(t)=\cdots=x_{l_1}(t),\ldots,x_{\sigma_N+1}(t)=\cdots=x_{L}(t)\}$. It is imposed frequently in the literature to result in cluster synchronization for various multi-agent systems, e.g., \cite{Yu_Wang2009group,Yu_Wang2010,Xia2011,Wu&Chen2009,Sun&Bai2011,Feng2014group,Qin&Yu2013,Yu&Qin2014}. To fulfill it, one can let positive and negative weights be balanced for all of the links directing from one cluster to any agent in another cluster. The negative weights for inter-cluster links is supposed to provide desynchronizing influences. Note also that with Assumption \ref{assump:zero_row_sums} each $L_{ii}$ is the Laplacian of a subgraph $\cG_i$, $i=1,\ldots,N$.

\emph{Notation}:  $\ones_{n}=[1,1,\ldots,1]^T\in \RR^{n}$. The identity matrix of dimension $n$ is
$I_n\in \RR^{n\times n}$. The symbol $blockdiag\{M_1,\ldots, M_N\}$ represents the block diagonal matrix constructed from the $N$ matrices $M_1,\ldots,M_N$. ``$\otimes$" stands for the Kronecker product. A symmetric positive (semi-) definite matrix $S$ is represented by $S>0(S\geq 0)$. $Re\lambda(A)$ is the real part of the eigenvalue of a square matrix $A$, and $\sigma(A)$ is the spectrum of $A$.

\section{Conditions for Achieving Cluster Synchronization} \label{sec:syn_leaderless}
In this section, we first present a necessary and sufficient algebraic clustering condition that entangles parameters from the Laplacian $\cL$ and the system matrices $A_i$'s. Then, we present some graph topological conditions which offer more intuitive interpretations. 

The following discussion makes use of the weighted graph Laplacian
\begin{equation}
\label{eq:CS_18}
\cL_c=\begin{bmatrix}
c_1 L_{11}&\cdots& L_{1N}\\
\vdots&\ddots&\vdots& \\
L_{N1}&\cdots&c_N  L_{NN}\end{bmatrix}\in\RR^{L\times L},
\end{equation}
and the following matrix:
\begin{equation}
\label{eq:CS_19}
\hat \cL_c=\begin{bmatrix}
c_1\hat L_{11}&\cdots&\hat L_{1N}\\
\vdots&\ddots&\vdots& \\
\hat L_{N1}&\cdots&c_N \hat L_{NN}\end{bmatrix}\in\RR^{(L-N)\times(L-N)},
\end{equation}
where each $\hat L_{ij}$, $i,j=1,\ldots,N$ is a block matrix defined as
\begin{equation}
\label{eq:CS_18c}
\hat L_{ij}=\tilde L_{ij}-\ones_{l_i}\gamma_{ij}^T,
\end{equation}
with
\begin{equation} 
\begin{split}
\gamma_{ij}&=[b_{\sigma_i+1,\sigma_j+2},\cdots,b_{\sigma_i+1,\sigma_j+l_j}]^T\in\RR^{l_j-1},\notag\\
\tilde L_{ij}&=\begin{bmatrix}
b_{\sigma_i+2,\sigma_j+2}&\cdots&b_{\sigma_i+2,\sigma_j+l_j}\\
\vdots&\ddots&\vdots \\
b_{\sigma_i+l_i,\sigma_j+2}&\cdots&b_{\sigma_i+l_i,\sigma_j+l_j}\\
\end{bmatrix}\in\RR^{(l_i-1)\times (l_j-1)}.\notag\\
\end{split}
\end{equation}
The two matrices $\cL_c$ and $\hat\cL_c$ have the following algebraic relationship.
\begin{lemma} \label{thm:lapalacian_reduce}
Under Assumption \ref{assump:zero_row_sums}, each diagonal block $L_{ii}$ in $\cL_c$ has exactly one zero eigenvalue if and only if the corresponding matrix $\hat L_{ii}$ defined in \eqref{eq:CS_18c} is nonsingular. Moreover, $\cL_c$ defined in \eqref{eq:CS_18} has exactly $N$ zero eigenvalues if and only if the matrix $\hat \cL_c$ defined in \eqref{eq:CS_19} is nonsingular.
\end{lemma}
The proof of this lemma is shown in Appendix \ref{appendix:proof_lapalacian_reduce}. This conclusion will be used frequently for deriving the main results in the following two subsections.
\subsection{Algebraic clustering conditions}
Under Assumption \ref{assump:stabilizable}, for each $i=1,\ldots,N$ there exists a matrix $P_i>0$ satisfying the Riccati equation
\begin{equation} \label{eq:CS_11}
P_iA_i+A_i^TP_i-P_iB_iB_i^TP_i=-I.\;\;
\end{equation}
Choose the control gain matrices as $K_i=-B_i^TP_i$, and denote $$\hat \bA=blockdiag\{I_{l_1-1}\otimes A_1,\ldots,I_{l_N-1}\otimes A_N\}.$$ Then, we have the following algebraic condition to check the cluster synchronizability of a linear multi-agent system.
\begin{theorem}\label{thm:clustersyn_leaderless}
Under Assumptions \ref{assump:stabilizable} to \ref{assump:zero_row_sums}, the multi-agent system in \eqref{sys:linearmodel} with couplings in \eqref{sys:controllaws_leaderless} achieves $N$-cluster synchronization if and only if
the matrix $\hat \bA-c \hat \cL_c\otimes I_n$ is Hurwitz, where $\hat \cL_c$ is defined in \eqref{eq:CS_19}.
\end{theorem}

The proof is given in Appendix \ref{appendix:proof_clustersyn_leaderless}. The matrix $\hat \bA-c \hat \cL_c\otimes I_n$ contains parameters from the interaction graph that entangle intimately with those from the system dynamics. In general, it is not possible to verify the above synchronization condition by simply comparing the eigenvalues of $\hat\cL$ with those of $A_i$'s. However, one can do so for a homogeneous multi-agent system as stated in the following corollary.

\begin{corollary}\label{thm:clustersyn_leaderless_hom}
Under Assumptions \ref{assump:stabilizable} to \ref{assump:zero_row_sums}, and with identical system parameters: $A_i=A$, $B_i=B$, $K_i=K,$ for all $i=1,\dots,N$, a multi-agent system in \eqref{sys:linearmodel} with couplings in \eqref{sys:controllaws_leaderless} achieves $N$-cluster synchronization if and only if the following holds:
\begin{equation} \label{eq:CS_22b}
\min_{\sigma(\hat\cL_c)} Re \lambda(c\hat\cL_c)>\max_{\sigma(A)} Re\lambda(A).
\end{equation}
\end{corollary}

A sketch of the proof for this corollary is given in Appendix \ref{appendix:proof_clustersyn_leaderless_hom}.
\begin{remark}
In words, the algebraic condition \eqref{eq:CS_22b} states that the weighted graph Laplacian $\cL_c$ has exactly $N$ zero eigenvalues, and all the nonzero eigenvalues have large enough positive real parts to dominate the unstable system dynamics described by $A$.
This condition implies that related results in \cite{Yu_Wang2009group,Yu_Wang2010,Xia2011} are special cases with $A=0$, $B=1$ and $K=1$. It also includes part of the results in \cite{Feng2014group}, which are obtained for identical double integrators. Note that with identical system parameters, one can use static controllers without involving the auxiliary variables $\eta_l$'s. However, in that case the synchronized state in each cluster depends linearly on the initials states $x_l(0)$'s only. For certain initial state sets, state separations in the limit cannot be guaranteed.
\end{remark}

\subsection{Graph topological conditions}\label{sec:topology_conditions_leaderless}

The matrix $\hat \bA-c \hat \cL_c\otimes I_n$ in Theorem \ref{thm:clustersyn_leaderless} can be proven to be Hurwitz for certain graph topologies in conjunction with some lower bounds on the weighting factors. To do so, the following well-known result for subgraphs will be useful.

\begin{lemma} [\cite{Ren2005}] \label{thm:eign_subgraphs_leaderless}
 Let $\cG_i$ be a non-negatively weighted subgraph. Then, the Laplacian $L_{ii}$ of $\cG_i$ has a simple zero eigenvalue and all the nonzero eigenvalues have positive real parts if and only if $\cG_i$ contains a directed spanning tree.
\end{lemma}

If a subgraph $\cG_i$ satisfies the conditions in Lemma \ref{thm:eign_subgraphs_leaderless}, then, by Lemma \ref{thm:lapalacian_reduce}, there exists a positive definite matrix $\hat W_i\in\RR^{(l_i-1)\times(l_i-1)}$ such that
\begin{equation}\label{eq:CS_12}
\hat W_i\hat L_{ii}+\hat L^T_{ii} \hat W_i>0,\;\; i=1,\ldots,N.
\end{equation}
Denote $$\hat \cW=blockdiag\{\hat W_1,\ldots,\hat W_N\},$$
and let
\begin{equation}
\hat \cL_o=\hat \cL_c-\hat \cL_d,
\end{equation}
with $\hat \cL_d=blockdiag\{c_1\hat L_{11},\ldots,c_N \hat L_{NN}\}$.
The following theorem states the main result of this subsection.
\begin{theorem}\label{thm:clustersyn_leaderless2}
Under Assumptions \ref{assump:stabilizable} to \ref{assump:zero_row_sums}, a multi-agent system in \eqref{sys:linearmodel} with couplings in \eqref{sys:controllaws_leaderless} achieves $N$-cluster synchronization exponentially fast 
if each subgraph, $\cG_i$, contains only cooperative edges and has a directed spanning tree, and the weighting factors satisfy
\begin{equation} \label{eq:CS_29}
c>\max_{i\in\{1,\ldots,N\}}\lambda_{\max}(A_i+A_i^T),
\end{equation}
and for each $i=1,\ldots,N$
\begin{equation} \label{eq:CS_20}
c_i\geq\dfrac{\lambda_{\max}(\hat \cW)-\lambda_{\min}(\hat \cW\hat \cL_o+\hat \cL_o^T \hat \cW)}{\lambda_{\min}(\hat W_i\hat L_{ii}+\hat L^T _{ii} \hat W_i)},
\end{equation}
where each $\hat W_i$ satisfies \eqref{eq:CS_12}.
\end{theorem}

\begin{proof}
Following the proof of the sufficiency part of Theorem \ref{thm:clustersyn_leaderless}, we need to show that the system
\begin{equation} \label{eq:CS_21}
\dot \zeta(t)=(\hat \bA-c \hat \cL_c\otimes I_n)
\zeta(t)\end{equation}
 is exponentially stable under the conditions in Theorem \ref{thm:clustersyn_leaderless2}. First, these conditions guarantee the existence of positive definite matrices, $\hat W_i$'s, satisfying \eqref{eq:CS_12}. Hence, \eqref{eq:CS_20} can be written as
\[c_i\lambda_{\min}( \hat W_i\hat L_{ii}+\hat L^T_{ii} \hat W_i)+\lambda_{\min}(\hat \cW \hat \cL_o+\hat \cL^T_o \hat \cW)\ge\lambda_{\max}(\hat \cW)\]
for $i=1,\ldots,N$. These inequalities together with Weyl's eigenvalue theorem (\cite{Horn&Johnson1987}) yield the following:
 \begin{align*}
  &\;\quad\lambda_{\min}(\hat \cW\hat \cL_c+\hat \cL^T_c \hat \cW)\\
  &=\lambda_{\min}(\hat \cW \hat \cL_d+\hat \cL^T_d \hat \cW+\hat \cW\hat \cL_o+\hat \cL^T_o \hat \cW)\\
  &\geq\lambda_{\min}(\hat \cW \hat \cL_d+\hat \cL^T_d \hat \cW)+\lambda_{\min}(\hat \cW\hat \cL_o+\hat \cL^T_o \hat \cW)\\
  &\geq\lambda_{\max}(\hat \cW),
 \end{align*}
which further implies that
\begin{equation} \label{eq:CS_22}
\hat \cW \hat \cL_c+\hat \cL^T_c \hat \cW \geq \hat \cW.
\end{equation}
Now, consider the Lyapunov function candidate $V(t)=\zeta(t)^T (\hat \cW\otimes I_n)\zeta(t)$ for the system \eqref{eq:CS_21}. Taking time derivative on both sides of $V(t)$, one gets
\begin{align*}
\dot V(t)&=\zeta^T(t)[ (\hat \cW\otimes I_n)(\hat \bA-c \hat \cL_c\otimes I_n)+(\hat \bA-c \hat \cL_c\otimes I_n)^T(\hat \cW\otimes I_n)] \zeta(t) \\
&=\zeta^T(t)[(\hat \cW\otimes I_n)(\hat \bA+\hat \bA^T)-c(\hat \cW\hat \cL_c+\hat \cL^T_c \hat \cW)\otimes I_n]\zeta(t)\\
&\leq \zeta^T(t)[(\hat \cW\otimes I_n)(\hat \bA+\hat \bA^T)-c\hat \cW\otimes I_n]\zeta(t)\\
&\leq \zeta^T(t)[(\hat \cW\otimes I_n)(\lambda_{\max}(\hat \bA+\hat \bA^T)-c)]\zeta(t)\\
&=-[c-\lambda_{\max}(\hat \bA+\hat \bA^T)] V(t),
\end{align*}
where the first inequality follows from \eqref{eq:CS_22}. Since $c-\lambda_{\max}(\hat \bA+\hat \bA^T)>0$ according to \eqref{eq:CS_29}, the exponential stability of system \eqref{eq:CS_21} is validated. 
\end{proof}

We have the following comments on the condition in \eqref{eq:CS_29}:
\begin{enumerate}
  \item From the above proof, one can find another lower bound for $c$ as follows:
      \begin{equation} \label{eq:CS_29b}
       c>\dfrac{\lambda_{\max}((\hat \cW\otimes I_n)(\hat \bA+\hat \bA^T))}{\lambda_{\min}(\hat \cW\hat \cL_c+\hat \cL^T_c \hat \cW)}.
      \end{equation}
      This bound is tighter than that in \eqref{eq:CS_29} since the inequality in \eqref{eq:CS_22} and $\lambda_{\max}(\hat W_i)>0$, $\lambda_{\max}(A_i+ A_i^T)\geq 0$ for any $i$ imply that the right-hand side (RHS) of \eqref{eq:CS_29b} $\leq \dfrac{\lambda_{\max}(\hat \cW)\lambda_{\max}(\hat \bA+\hat \bA^T)}{\lambda_{\max}(\hat \cW)}=$ RHS of \eqref{eq:CS_29}. However, the tighter bound \eqref{eq:CS_29b} only guarantees that $\dot V(t)<0$, and does not specify the convergence rate. Moreover, the RHS of \eqref{eq:CS_29b} involves all the $c_i$'s in $\hat\cL_c$, and no known distributed algorithm is available for the computation.
  \item Note that the role of $c$ is more essential in stabilizing the unstable modes of the system matrices, $A_i$'s, than in strengthening the connective ability of the interaction graph. A global weighting factor similar to $c$ is utilized in a related paper \cite{Yu&Qin2014} where the clustering problem for identical linear systems are solved via a pinning control approach. However, the global factor in \cite{Yu&Qin2014} serves as a parameter in a Ricatti inequality so as to result in a control gain matrix. 
      In contrast, the selection of $c$ in this paper is independent of the control designs in \eqref{eq:CS_11}. 
\end{enumerate}

The following two remarks explain why the commonly used static couplings are not suitable choices when dealing with nonidentical linear systems.
\begin{remark} \label{rem:comparisons to identical}
To achieve state cluster synchronization for a group of generic linear systems, a natural choice of static couplings is the following  (slightly modified from the static couplings for homogeneous linear systems in \cite{Qin&Yu2013,Yu&Qin2014}):  for each $l\in\cC_i$, $i=1,\ldots,N$
  \begin{align}\label{sys:controllaws_static}
    &u_l(t)=K_i \left[c_i\sum_{k\in\cC_i}b_{lk}x_k(t)+\sum_{k\notin\cC_i}b_{lk}x_k(t)\right]
  \end{align}
However, following a similar procedure as in \cite{Yu&Qin2014}, one will need the following condition
  \begin{equation} \label{eq:CS_16}
c_i\lambda_{\min}((\hat W_i\hat L_{ii}+\hat L^T _{ii} \hat W_i)\otimes P_iB_iB^T_iP_i)\geq\rho,
\end{equation}
for each $i=1,\ldots,N$, where $\rho=\lambda_{\max}((\hat\cW\otimes I_n) \bP\bB\bB^T\bP)-\lambda_{\min}(\bP\bB\bB^T\bP(\hat\cW \cL_o\otimes I_n)+ (\cL_o^T\hat\cW\otimes I_n)\bP\bB\bB^T\bP)$. To compute $\rho$, one needs information on the control design of all agents, i.e., $\bB^T\bP=blockdiag\{I_{l_1-1}\otimes B_1P_1,\ldots,I_{l_N-1}\otimes B_NP_N\}$. This fact renders the selection of local weighting factors, $c_i$'s, a centralized decision. Moreover, \eqref{eq:CS_16} cannot be satisfied by any $c_i$ in the nontrivial case that $\rho>0$ and $P_iB_iB^T_iP_i$ is singular for some $i$.
In contrast to \eqref{eq:CS_16}, the condition \eqref{eq:CS_20} specifies explicitly the requirements for $c_i$'s, and it is independent of the design of control gain matrices. In this sense, the dynamic couplings in \eqref{sys:controllaws_leaderless} are preferable to the static ones in \eqref{sys:controllaws_static}.
\end{remark}
\begin{remark}
For nonidentical nonlinear systems of the form, $\dot x_l(t)=f_i(x_l,t)$, $l\in\cC_i$, static couplings are used to result in closed-loop systems as follows (\cite{Sun&Bai2011,Lu&Chen2010}):
\begin{align*}
&\dot x_l(t)=f_i(x_l,t)- \Gamma\left[c_i\sum_{k\in\cC_i}b_{lk}x_k(t)+\sum_{k\notin\cC_i}b_{lk}x_k(t)\right],
\end{align*}
where $\Gamma$ is a constant (usually nonnegative-definite) matrix. 
It was shown that clustering conditions involve only the graph Laplacian (see \cite{Lu&Chen2010}) if all individual self-dynamics are constrained by the so-called QUAD condition: for any $x,y\in\RR^n$, $ (x-y)^T[f_l(x)-f_l(y)- \Gamma(x-y)]\leq -\omega (x-y)^T(x-y)$, where $\omega>0$ is a prescribed positive scalar. For generic linear systems with static couplings in \eqref{sys:controllaws_static}, this QUAD condition requires that for any $x\in\RR^n$, $x^T(A_i- \Gamma)x\leq -\omega x^T x$ with $\Gamma=B_iK_i$ for all $i=1,\ldots,N$. Given a $\Gamma$, for the existence of control gains $K_i$'s, one needs all $B_i$'s to satisfy $Rank(B_i)=Rank([B_i\; \Gamma])$. However, this rank condition is too restrictive. For example, for the models in \eqref{eq:ship_matrix}, an applicable choice of $\Gamma$ is $I_2$, but then $Rank(B_i)<Rank([B_i\; \Gamma])$ and thus no $K_i$ can be solved from $\Gamma=B_iK_i$. In contrast, the dynamic couplings in \eqref{sys:controllaws_leaderless} do not impose such constraints on the system models.
\end{remark}

Generally, it is not always necessary to let every subgraph contain a directed spanning tree. In fact, agents belonging to a common cluster may not need to have direct connections at all as long as the algebraic condition in Theorem \ref{thm:clustersyn_leaderless} is satisfied. This point is illustrated by a simulation example in the next section.   Nevertheless, the spanning tree condition turns out to be necessary under some particular graph topologies as stated by the corollary below.

\begin{corollary}\label{thm:clustersyn_leaderless_acyclic}
Let $\cG$ be an interaction graph with an acyclic partition as in \eqref{var:laplacianpartition_triang}, and let the edge weights of every subgraph $\cG_i$ be nonnegative. Under Assumptions \ref{assump:stabilizable} to \ref{assump:zero_row_sums}, a multi-agent system \eqref{sys:linearmodel} with couplings in \eqref{sys:controllaws_leaderless} achieves $N$-cluster synchronization  if and only if every $\cG_i$ contains a directed spanning tree, and the weighting factors satisfy
\begin{equation} \label{eq:CS_31}
c\cdot c_i>\dfrac{\max_{\sigma(A_i)} Re\lambda(A_i)}{\min_{\sigma(\hat L_{ii})} Re\lambda (\hat L_{ii})},\;\forall i=1,\ldots,N,
\end{equation}
where each $\hat L_{ii}$ is defined in \eqref{eq:CS_18c}.
\end{corollary}
\begin{proof}
By Theorem \ref{thm:clustersyn_leaderless}, we can examine the stability of $\hat \bA-c\hat \cL_c\otimes I_n$.
 Let $T_i\in\RR^{(l_i-1)\times (l_i-1)}$, $i=1,\ldots,N$, be a set of nonsingular matrices such that $ T_i^{-1}\hat L_{ii} T_i= J_i,$
where $J_i$ is the Jordan form of $\hat L_{ii}$.
Denote $\bT=blockdiag\{T_1\otimes I_n,\ldots,T_N\otimes I_n\}$. Then, the block triangular matrix $\bT^{-1}(\hat \bA-c\hat \cL_c\otimes I_n)\bT$ has diagonal blocks $A_{i}-\tilde c_i\lambda_k ( \hat L_{ii})I_n$, where $\tilde c_i=c\cdot c_i$, $k=1,\ldots,l_i-1$, $i=1,\ldots,N$. Hence, the matrix $\hat \bA-c\hat \cL_c\otimes I_n$ is Hurwitz if and only if  $\tilde c_i \min_{k} Re \lambda_k (\hat L_{ii})>\max_{m} Re \lambda_m (A_{i})$ for any $i$. This claim is equivalent to the conclusion of this corollary due to Lemma \ref{thm:eign_subgraphs_leaderless}, the first claim of Lemma \ref{thm:lapalacian_reduce}, and Assumption \ref{assump:A_unstable} that requires $\max_{m} Re \lambda_m (A_{i})\geq 0$.
\end{proof}

This corollary reveals the indispensability of \emph{direct} links among agents in the same cluster under an acyclically partitioned interaction graph. 
Note that such direct interaction requirement for intra-cluster agents is not necessary under a nonnegatively weighted interaction graph (see \cite{Lu&Chen2010,Han&Chen2013,Han&Chen2015} for references).
\begin{remark}It is worth mentioning for the condition in \eqref{eq:CS_31} that one can set $c_i=1$ for all $i$, and adjust the global factor $c$ only to result in cluster synchronization. In contrast, without the acyclic partitioning structure, the local weighting factors $c_i$'s need to satisfy the lower bound conditions in \eqref{eq:CS_20}. Note that \eqref{eq:CS_31} specifies the tightest lower bound for $c$, while a lower bound reported in  \cite{Qin&Yu2013} for identical linear systems via Lyapunov stability analysis can be quite loose. 
\end{remark}

\section{Applications and Simulation Examples} \label{sec:simulation}
In this section, we provide application examples for cluster synchronization of nonidentical linear systems. We also conduct numerical simulations using these models to illustrate the derived theoretical results.
\subsection{Example 1: Heading alignment of nonidentical ships}

\begin{figure}[h]
  \centering
  \subfigure[Interaction graph partitioned into two clusters $\cC_1=\{1,2\}$ and $\cC_2=\{3,4\}$.]{
    \includegraphics[width=5.5cm] {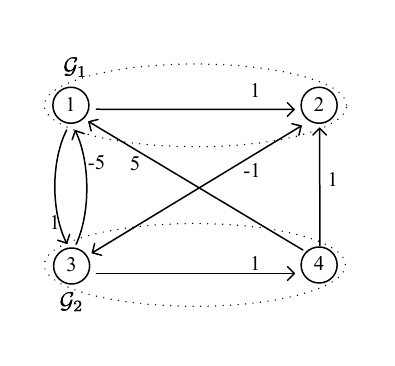}
  \label{fig:topology_4nodes_a1}
  }  \hspace*{5mm}
  \subfigure[The heading angles $\psi_l(t)$ synchronize into two groups.]{
    \includegraphics[width=6.8cm] {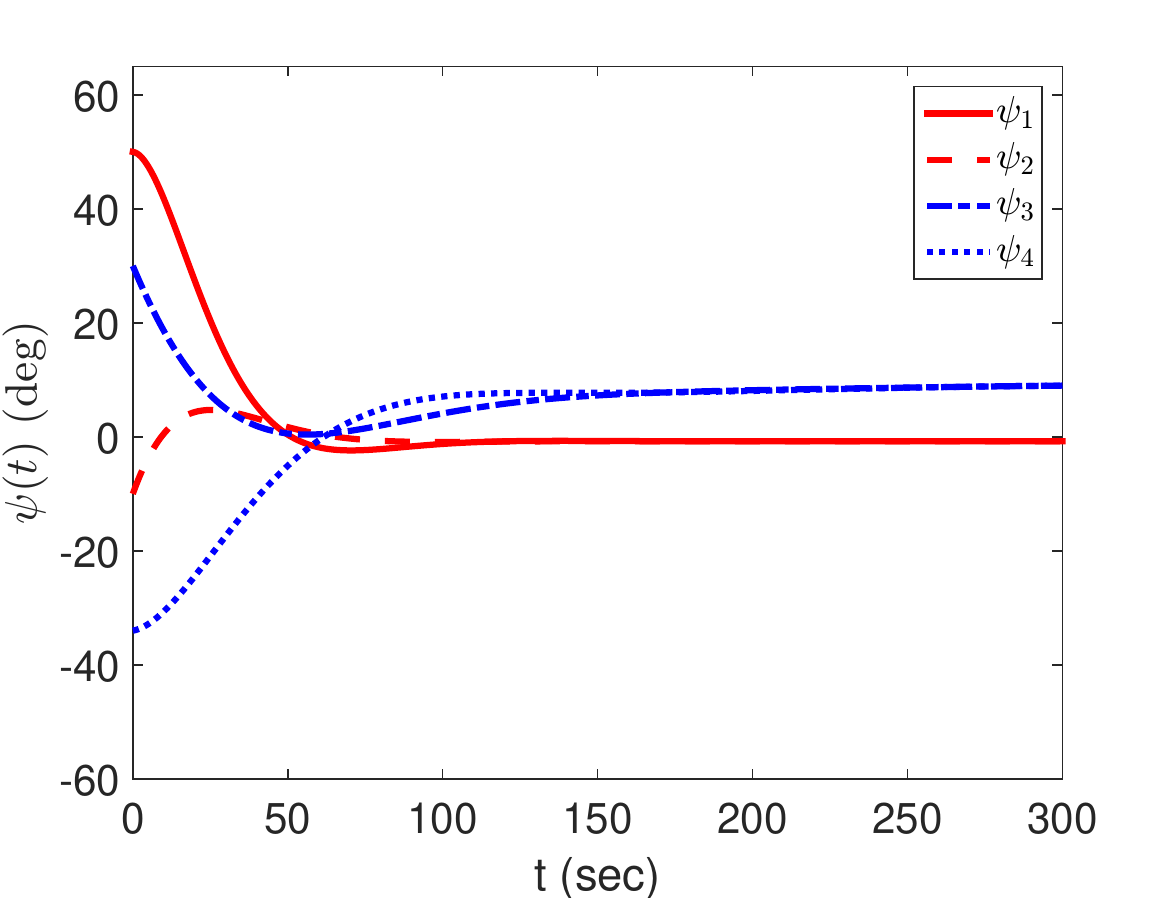}
  \label{fig:clustering_ship1}
  }

  \subfigure[The velocities $v_l(t)$ of all ships converge to zero.]{
    \includegraphics[width=6.8cm] {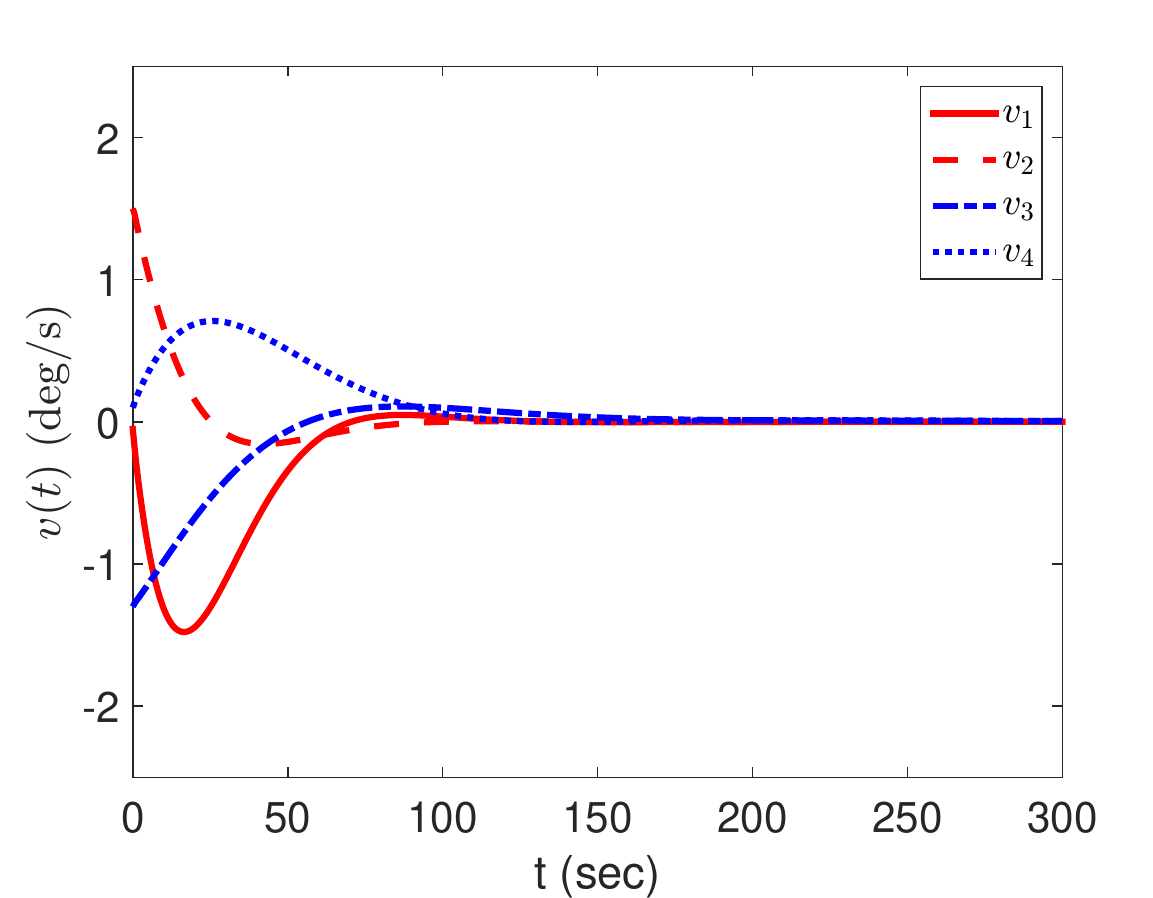}
  \label{fig:clustering_ship1_velocity}
  }
  \hspace*{1mm}
  \subfigure[All auxiliary control variables $\eta_l(t)$ converge to zero.]{
    \includegraphics[width=6.8cm] {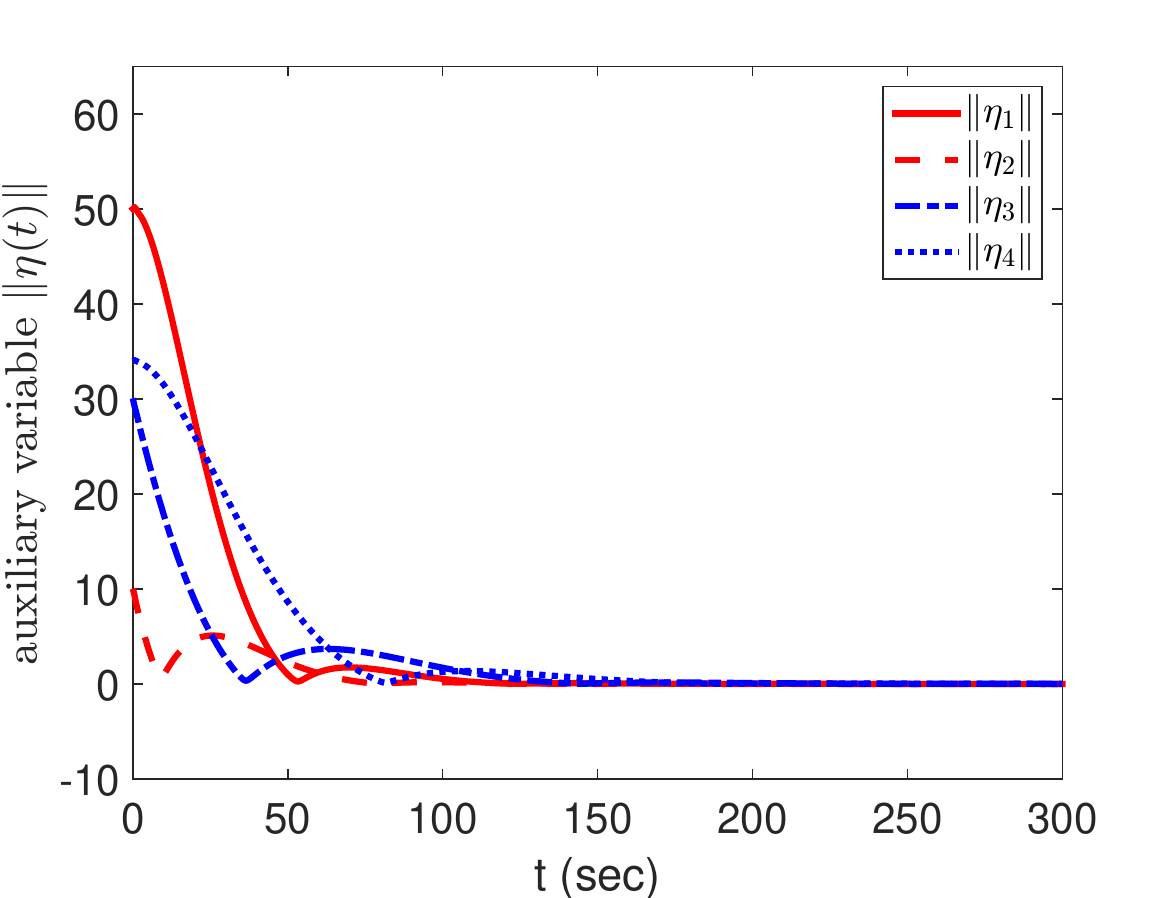}
  \label{fig:clustering_ship1_aux}
  }
  \caption{Cluster synchronization for systems in \eqref{eq:ship_model} under interaction graph in Figure \ref{fig:topology_4nodes_a1}.}
\end{figure}

Consider a group of four ships with the interaction graph described by Figure \ref{fig:topology_4nodes_a1}, where ship 1 and 2 (respectively, ship 3 and 4) are of the same type. The purpose is to synchronize the heading angles for ships of the same type. The steering dynamics of a ship is described by the well-known Nomoto model \cite{Fosseng1994}:
\begin{align} \label{eq:ship_model}
\dot\psi_l(t)&=v_l(t)\notag\\
\dot v_l(t)&=-\dfrac{1}{\tau_i} v_l(t)+\dfrac{\kappa_i}{\tau_i} u_l(t)
\end{align}
where $\psi_l$ is the heading angle (in degree) of a ship $l\in\cI$, $v_l$ (deg/s) is the yaw rate, and $u_l$ is the output of the actuator (e.g., the rudder angle). The parameter $\tau_i$ is a time constant, and $\kappa_i$ is the actuator gain, both of which are related to the type of a ship.

Define the system matrices
\begin{equation} \label{eq:ship_matrix}
A_i=\begin{bmatrix}
0&1\\
0&-\frac{1}{\tau_i}\\
\end{bmatrix},\;\;
B_i=\begin{bmatrix}
0\\
\frac{\kappa_i}{\tau_i}
\end{bmatrix},
\end{equation}
for $i=1,2$, and assume that $\tau_1=42.21$, $\tau_2=107.3$, $\kappa_1=0.181$, $\kappa_2=0.185$. Clearly, these system matrices satisfy Assumptions \ref{assump:stabilizable} $\&$ \ref{assump:A_unstable}. The solutions to the Riccati equations in \eqref{eq:CS_11} are given by
$P_1 =\begin{bmatrix}
    22.3   & 233.2\\
    233.2  & 3915.4
    \end{bmatrix}$ and
$P_2 =\begin{bmatrix}
     34  & 580\\
    580   &16875
    \end{bmatrix}$, which lead to the control gain matrices $K_1=-[1\;\; 16.79]$ and $K_2=-[1\;\; 29.09]$.

The weighted graph Laplacian of the interaction graph in Figure \ref{fig:topology_4nodes_a1} is given by
\begin{equation*}
\cL_c=\begin{bmatrix}
0&0&5&-5\\
-c_1&c_1&1&-1\\
-1&1&0&0\\
0&0&-c_2&c_2\\
\end{bmatrix},
\end{equation*}
which satisfies Assumption \ref{assump:zero_row_sums} when partitioned from the second row and column with respect to $\cC_1=\{1,2\}$ and $\cC_2=\{3,4\}$. Using the definition in \eqref{eq:CS_19} yields $\hat \cL_c=\begin{bmatrix}
c_1&4\\
-1&c_2\\
\end{bmatrix}$, which indicates that $\hat L_{11}=1$ and $\hat L_{22}=1$. Hence, the inequalities in \eqref{eq:CS_12} hold for any $\hat W_1>0$ and $\hat W_2>0$. We choose $\hat W_1=\hat W_2=1$. It follows that $\lambda_{\max}(\hat \cW)=1$, $\lambda_{\min}(\hat \cW\hat \cL_o+\hat \cL_o^T \hat \cW)=-3$, and $\lambda_{\min}(\hat W_i\hat L_{ii}+\hat L^T _{ii} \hat W_i)=2$ for $i=1,2$. Then, we can choose $c_1=c_2=2$ so that the inequalities in \eqref{eq:CS_20} are satisfied. Since $\max_{i=1,2}\lambda_{\max}(A_i+A_i^T)=0.99,$  we set $c=1$ according to \eqref{eq:CS_29}. Further noticing that the subgraph of each cluster in Figure \ref{fig:topology_4nodes_a1} contains a spanning tree and the edges in each subgraph all have positive weights, we see that all conditions in Theorem \ref{thm:clustersyn_leaderless2} are met. Note also that with the parameters designed above, the matrix $\hat \bA- c\hat\cL_c\otimes I_2$ in Theorem \ref{thm:clustersyn_leaderless} has eigenvalues $\{-2 \pm 2j, -2.0165 \pm 2 j\}$ where $j=\sqrt{-1}$, and thus is Hurwitz. The simulation result in Figure \ref{fig:clustering_ship1} shows that cluster synchronization is achieved for the heading angles (the velocity $v_l(t)$ of every agent will be stabilized to zero as shown in Figure \ref{fig:clustering_ship1_velocity}), and the auxiliary control variables $\eta_l(t)$ of all ships converge to zero as shown in Figure \ref{fig:clustering_ship1_aux}. 

\begin{figure}[ht]
  \centering
  \subfigure[Interaction graph partitioned into two clusters where nodes in the first cluster have no direct interaction.]{
    \includegraphics[width=5.5cm] {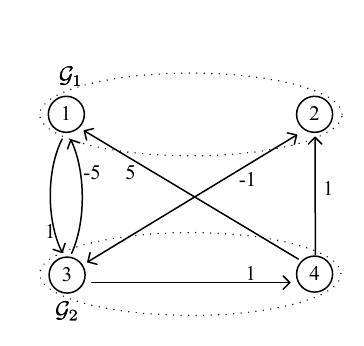}
  \label{fig:topology_4nodes_a2}
  }\hspace*{10mm}
  \subfigure[Under the interaction graph in Figure \ref{fig:topology_4nodes_a2}, the heading angles $\psi_l(t)$ of the ships synchronize into two groups.]{
    \includegraphics[width=7cm] {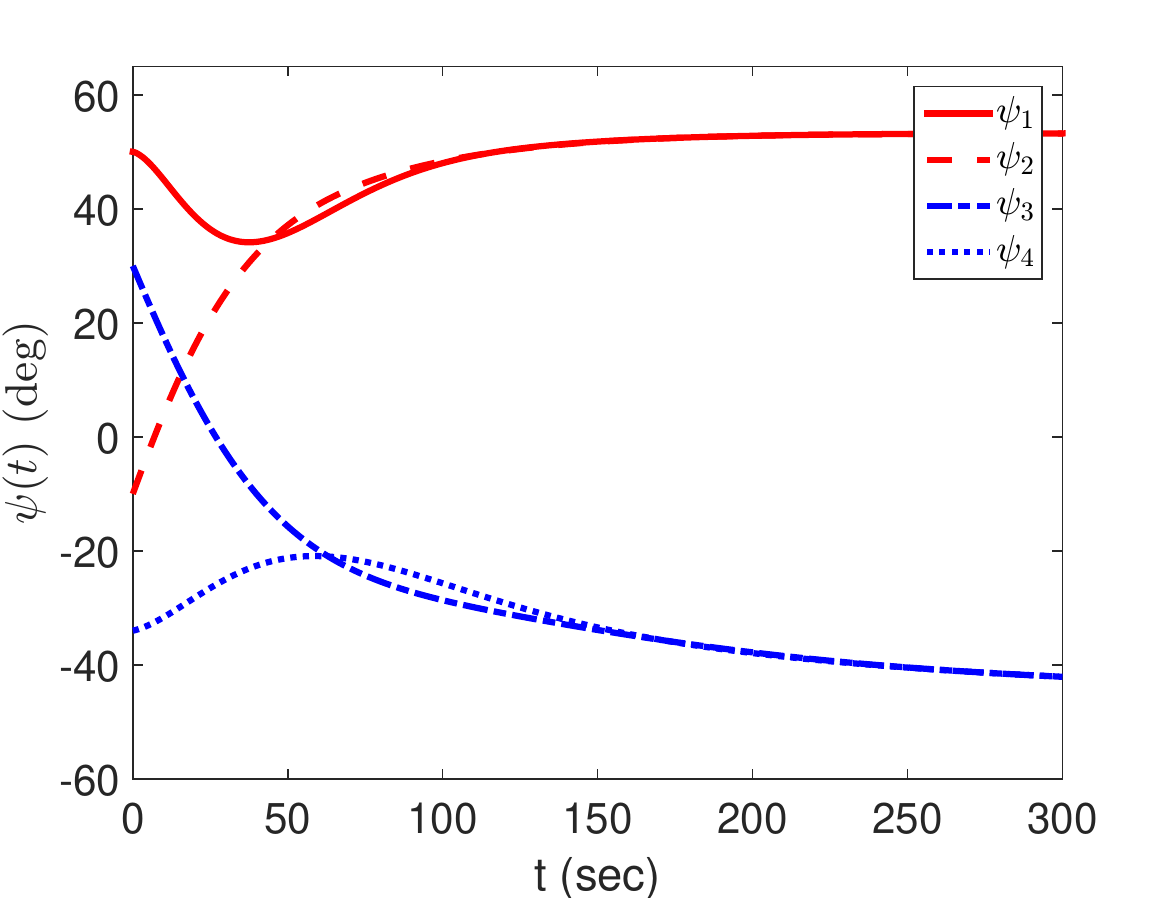}
  \label{fig:clustering_ship2}
  }\hspace*{1mm}
  \caption{Cluster synchronization is achieved for systems in \eqref{eq:ship_model} under the above interaction graph.}
\end{figure}

\begin{figure}[ht]
\centering
  \subfigure[Interaction graph partitioned \emph{acyclically} into two clusters where nodes in the first cluster have no direct interaction.]{
    \includegraphics[width=5cm] {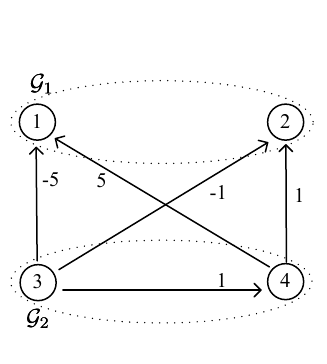}
  \label{fig:topology_4nodes_b}
  }\hspace*{15mm}
  \subfigure[The heading angles $\psi_1$ and $\psi_2$ of ships in cluster $\cC_1$ \emph{did not} synchronize together.]{
    \includegraphics[width=7cm] {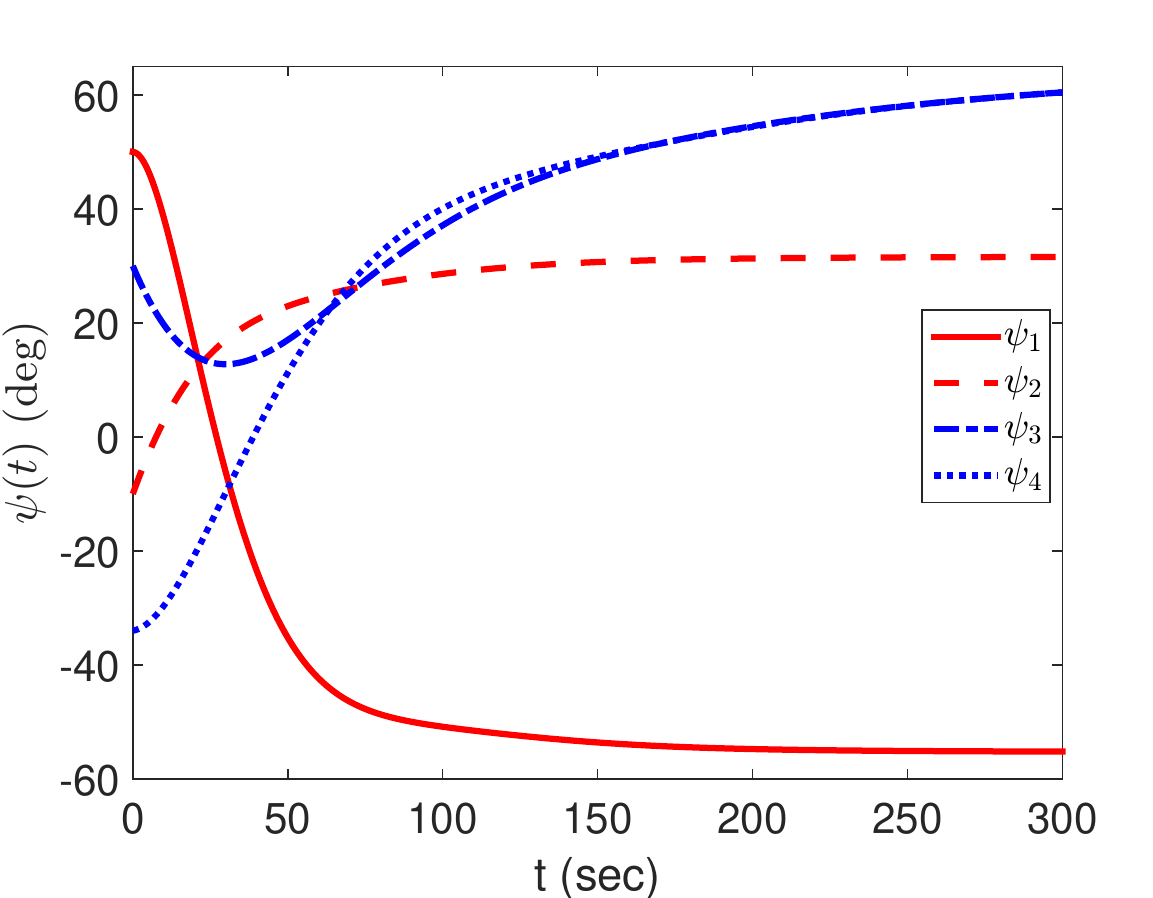}
  \label{fig:notclustering_shipacyclic}
  }
  \caption{Under an acyclically partitioned graph, the systems in \eqref{eq:ship_model} cannot achieve cluster synchronization.}
\end{figure}

Now, let $c_1=0$ so that agents $1$ and $2$ in cluster $\cC_1$ have no direct connection as shown in Figure \ref{fig:topology_4nodes_a2}. Then the matrix $\hat \bA- c\hat\cL_c\otimes I_2$ in Theorem \ref{thm:clustersyn_leaderless} has eigenvalues $\{  -1 \pm 1.7321i,  -1.0165 \pm 1.7362i \}$, and thus is still Hurtwiz. Simulation result in Figure \ref{fig:clustering_ship2} shows that cluster synchronization is achieved for the heading angles (The velocities $v_l(t)$ and auxiliary control variables $\eta_l(t)$, $l\in\cI$ all converge to zero in the simulation, and their evolution figures are omitted for simplicity). This example illustrates that containing a spanning tree for the subgraph of each cluster is only a sufficient condition for achieving cluster synchronization under a cyclically partitioned interaction graph.
However, with an acyclic partition as in Figure \ref{fig:topology_4nodes_b}, the agents in cluster $\cC_1$, having no direct connections, cannot achieve state synchronization as shown in Figure \ref{fig:notclustering_shipacyclic}. This verifies the necessity of the spanning tree condition in Corollary \ref{thm:clustersyn_leaderless_acyclic}. Furthermore, observing the following matrix associated with the acyclically partitioned graph in Figure \ref{fig:topology_4nodes_b}
$$\hat \bA- c\hat\cL_c\otimes I_2=
\begin{bmatrix}
0&1&&\\
0&-\frac{1}{\tau_1}&&\\
&&0&1\\
&&0&-\frac{1}{\tau_2}
\end{bmatrix}
-c\begin{bmatrix}
0&4I_2\\
0&c_2 I_2
\end{bmatrix},
$$
we find that it cannot be made stable for any $c>0$ and  $c_2>0$. So, this example also verifies the necessity of the algebraic condition in Theorem \ref{thm:clustersyn_leaderless}.


\subsection{Example 2: Cluster synchronization of harmonic oscillators}
\begin{figure} [ht]
\centering
  \subfigure[Interaction graph partitioned into two clusters $\cC_1=\{s_1, r_1, r_2\}$ and $\cC_2=\{s_2, r_3, r_4\}$.]{
  \includegraphics[width=5cm]{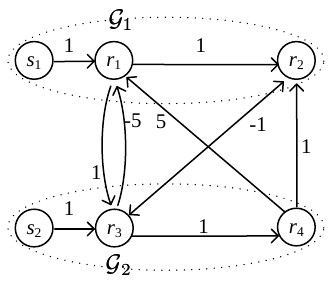}
  \label{fig:topology_6nodes}
  }\hspace{7mm}
  \subfigure[The first components $x_{1l}(t)$ in \eqref{eq:harmonic_osc} synchronize into two groups.]{
  \includegraphics[width=7cm] {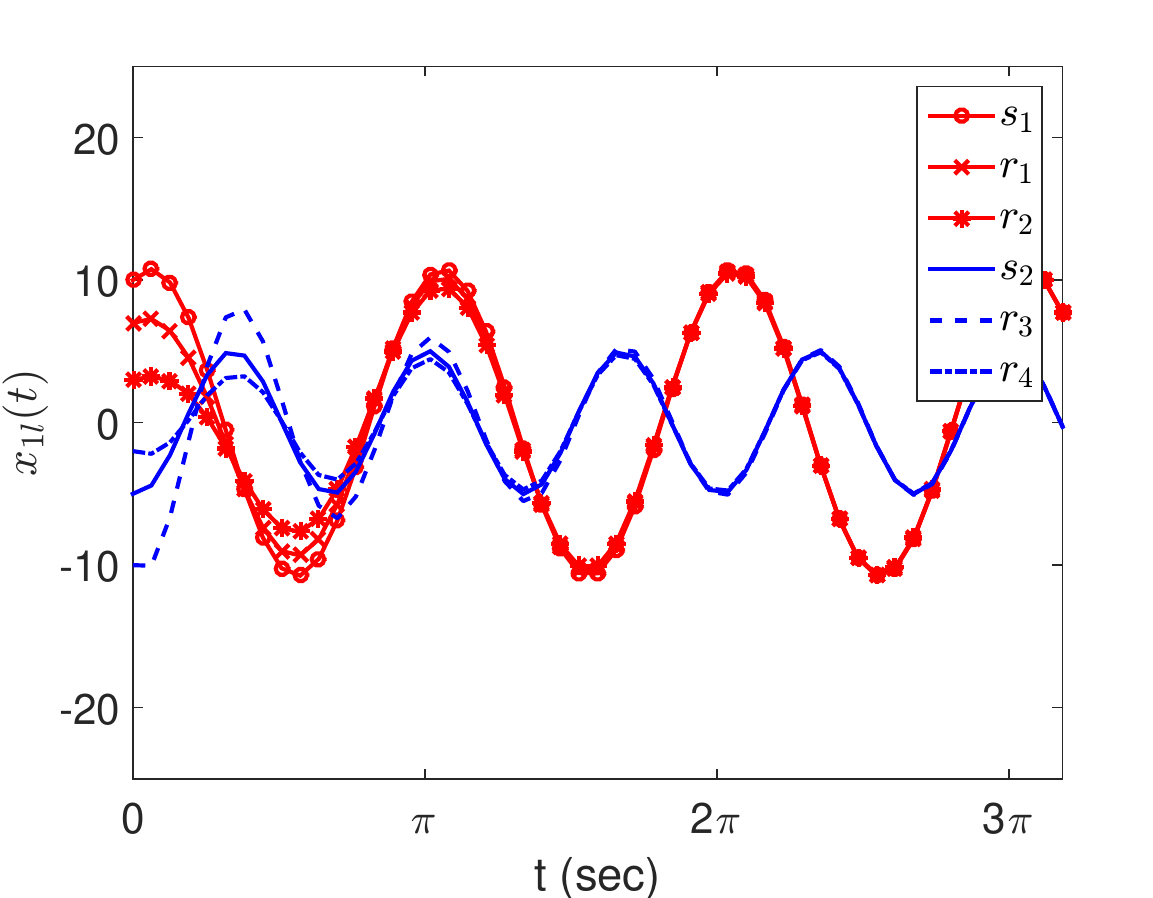}
  \label{fig:clustering_oscillator_state1}
  }
  \subfigure[The second components $x_{2l}(t)$ in \eqref{eq:harmonic_osc} synchronize into two groups.]{
  \includegraphics[width=6.8cm] {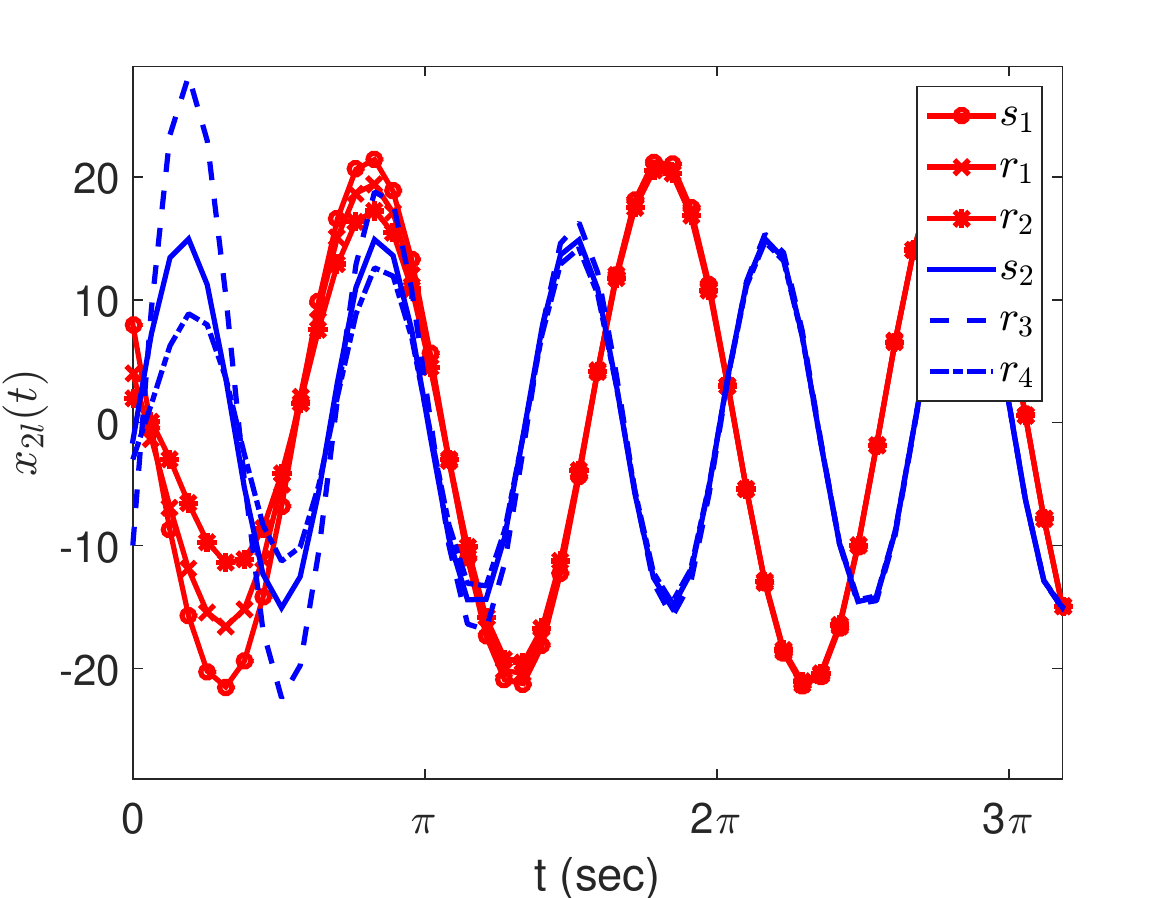}
  \label{fig:clustering_oscillator_state2}
  }
  \hspace*{1mm}
  \subfigure[The auxiliary control variables of all receivers converge to zero.]{
  \includegraphics[width=6.8cm] {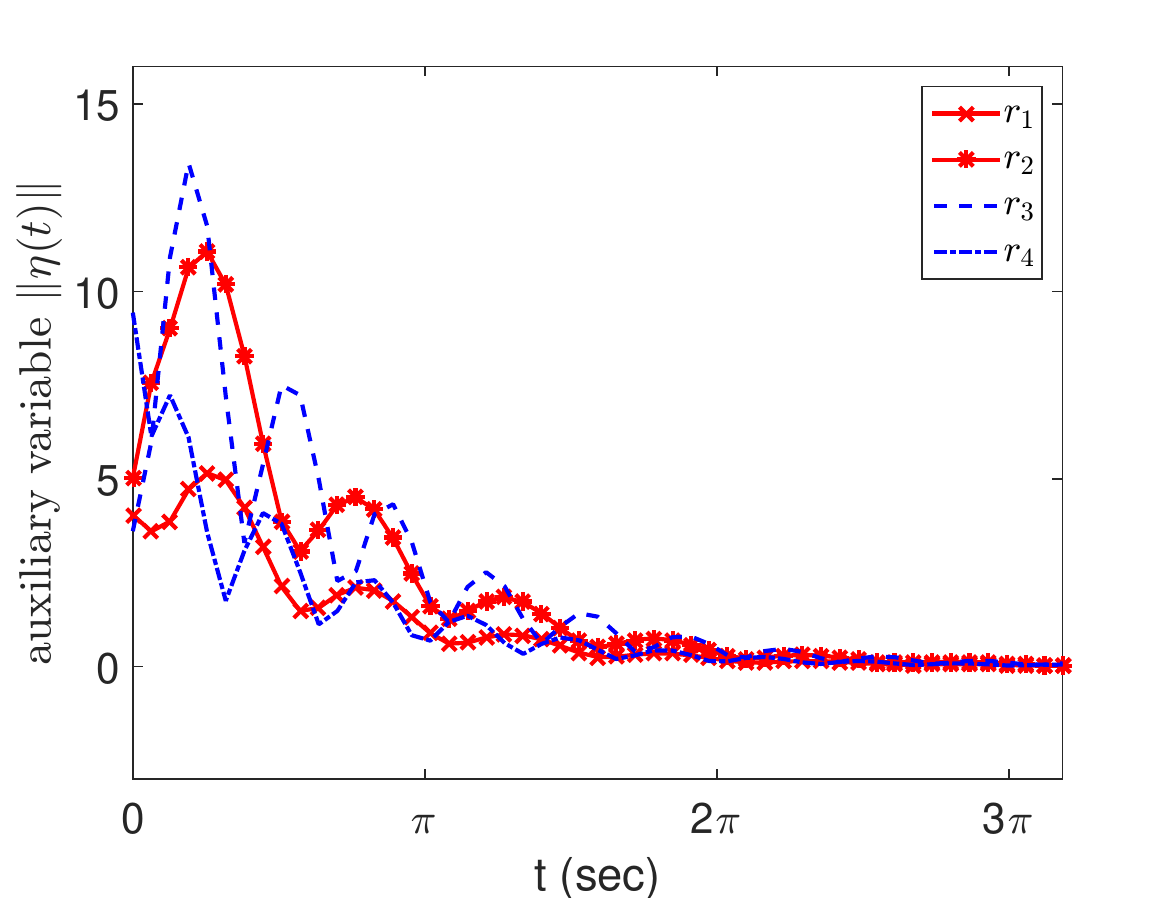}
  \label{fig:clustering_oscillatoraux}
  }
\caption{Cluster synchronization for the nonidentical harmonic oscillators in \eqref{eq:harmonic_osc}.}
\end{figure}

The studied cluster synchronization problem for nonidentical linear systems may find applications in the coexistence of oscillators with different frequencies. To see this, let us consider two clusters of coupled harmonic oscillators with graph topology in Figure \ref{fig:topology_6nodes}, where the first cluster contains a sender $s_1$ and two receivers $r_1$ and $r_2$, the second cluster contains a sender $s_2$ and two receivers $r_3$ and $r_4$, and the four receivers are coupled by some directed links. Assume the angular frequencies of the two clusters of oscillators are $w_1=2$ rad/s and $w_2=3$ rad/s, respectively. Then, the dynamic equation of each oscillator is given by (\cite{Ren2008})
\begin{align} \label{eq:harmonic_osc}
\dot x_{1l}(t)&=x_{2l}(t),\notag\\
\dot x_{2l}(t)&=-w_i^2x_{1l}(t)+u_l(t),\;l\in\cC_i,\;i=1,2
\end{align}
which corresponds to the following system matrices:
\begin{equation*}
A_i=\begin{bmatrix}
0&1\\
-w_i^2&0\\
\end{bmatrix},\;\;
B_i=\begin{bmatrix}
0\\
1
\end{bmatrix}, \;\;i=1,2.
\end{equation*}
The objective is to let the receivers of each cluster follow the state of the sender.

Note that the above system matrices satisfy Assumptions \ref{assump:stabilizable} $\&$ \ref{assump:A_unstable}, and the Laplacian matrix of the interaction graph in Figure \ref{fig:topology_6nodes} satisfies Assumption \ref{assump:zero_row_sums}. Besides, every subgraph $\cG_i,i=1,2$ in Figure \ref{fig:topology_6nodes} contains a directed spanning tree with the receiver being the root node. Then, following a similar design procedure as in the previous example, we can set $K_1 =-[0.1231\;1.1163]$, $K_2 = -[0.0554\; 1.0539]$, $c=6$, and $c_1=c_2=13$. Simulation results in Figure \ref{fig:clustering_oscillator_state1} and Figure \ref{fig:clustering_oscillator_state2} show the synchronized oscillations of the harmonic oscillators with two distinct frequencies. Figure \ref{fig:clustering_oscillatoraux} shows the convergence of all receivers' auxiliary control variables. (Note that the two senders don't need to be controlled. Hence their auxiliary control variables stay at zero all the time and are not shown in the figure). This example indicates that our result can include the leader-follower structure (or pinning control approach \cite{Yu&Qin2014}) as a special case since the senders play the role of leaders and the receivers can be considered as followers.


\section{Conclusions} \label{sec:conclusion}
This paper investigates the state cluster synchronization problem for multi-agent systems with nonidentical generic linear dynamics. 
By using a dynamic structure for coupling strategies, this paper derives both algebraic and graph topological clustering conditions which are independent of the control designs. 
For future studies, cluster synchronization which can only be achieved for the system outputs is a promising topic, especially for linear systems with parameter uncertainties or for heterogeneous nonlinear systems. For completely heterogenous linear systems, research works following this line are conducted by the authors in \cite{Liu&Wong_CDC2015} and others in \cite{Liu&Cao2014}. For nonlinear heterogeneous systems, the new theory being established for \emph{complete output} synchronization problems  \cite{Wieland2011,Su&Huang2013NROR} may be further extended. Another interesting challenge existing in cluster synchronization problems is to discover other graph topologies that meet the algebraic conditions. If the interactions among agents are based on communication systems, the issue of reducing communication demands by event-triggered control techniques \cite{Meng2016,Yang2016} would also be a promising topic for future studies.

\addtolength{\textheight}{0cm}   

\section*{Appendix}
\subsection{Proof of Lemma \ref{thm:lapalacian_reduce}} \label{appendix:proof_lapalacian_reduce}
\begin{proof}
Denote
$S_i=\begin{bmatrix}
1 &0\\
\ones_{l_i-1} &I_{l_i-1}
\end{bmatrix}\in\RR^{l_i\times l_i}$ for $i=1,\ldots,N$, and let $\bS=blockdiag\{S_1,\ldots,S_N\}$. Clearly, $S_i$ has the inverse matrix $S_i^{-1}=\begin{bmatrix}
1 &0\\
-\ones_{l_i-1} &I_{l_i-1}
\end{bmatrix}$.
By direct computation one can show that
$
S_i^{-1}L_{ij}S_j=\begin{bmatrix}
0&\gamma_{ij}\\
0&\hat L_{ij}
\end{bmatrix}.
$
This implies the first claim when $i=j$.

For the second claim, consider that
\[
\bS^{-1}\cL_c\bS=\begin{bmatrix}
0&\gamma_{11}&\cdots&0&\gamma_{1N}\\
0&c_1\hat L_{11}&\cdots&0&\hat L_{1N}\\
\vdots&\vdots&\ddots&\vdots&\vdots \\
0&\gamma_{N1}&\cdots&0&\gamma_{NN}\\
0&\hat L_{N1}&\cdots&0&c_N\hat L_{NN}\\
\end{bmatrix}.
\]
Rearrange the columns and rows of $\bS^{-1}\cL_c\bS$ by permutation and similarity transformations to get the following block upper-triangular matrix
\[
\begin{bmatrix}
0_{1\times N}&\gamma_{11}&\cdots&\gamma_{1N}\\
\vdots&\vdots&\ddots&\vdots\\
0_{1\times N}&\gamma_{N1}&\cdots&\gamma_{NN}\\
0_{(L-N)\times N}&&\hat \cL_c&\\
\end{bmatrix},
\]
where $\hat \cL_c$ is defined in \eqref{eq:CS_19}.
Then, the second claim of this lemma follows immediately.
\end{proof}
\subsection{Proof of Theorem \ref{thm:clustersyn_leaderless}} \label{appendix:proof_clustersyn_leaderless}
\begin{proof}
The closed-loop system equations for \eqref{sys:linearmodel} using couplings \eqref{sys:controllaws_leaderless} are given as
\begin{equation}\label{sys:linearmodel_leaderless}
\dot z_l=A_{ci} z_l-c\left[c_i\sum_{k\in\cC_i}b_{lk} E z_k+\sum_{k\notin\cC_i}b_{lk} E z_k \right],
\end{equation}
for all $l\in\cC_i,\; i=1\ldots,N$, where $z_l=[ x_l^T,\eta_l^T]^T$ and
\begin{equation} \label{eq:CS_7}
A_{ci}=\begin{bmatrix}
A_i & B_i K_i\\
0 & A_i+B_i K_i\\
\end{bmatrix},\;\;
E=\begin{bmatrix}
 0 &0\\
-I_n &I_n
\end{bmatrix}.
\end{equation}
Let $e_l(t):=z_l(t)-z_{\sigma_i+1}(t)$ for $l\in\cC_i$ and $l\neq \sigma_i+1$, $i=1,\ldots,N$.
It follows from \eqref{sys:linearmodel_leaderless} and Assumption \ref{assump:zero_row_sums} that
\begin{align}\label{eq:CS_3}
\dot e_l(t)&=A_{ci} e_l(t)-c\left[c_i\sum_{k\in\cC_i}(b_{lk}-b_{\sigma_i+1,k}) E e_k(t) +\sum_{k\notin\cC_i}(b_{lk}-b_{\sigma_i+1,k}) E e_k(t)\right].
\end{align}
Define a nonsingular transformation matrix $Q$ as follows:
\begin{equation} \label{eq:CS_8}
Q=\begin{bmatrix}
I_n &  0\\
I_n & I_n     \\
\end{bmatrix},\;\;
Q^{-1}=\begin{bmatrix}
I_n &  0\\
-I_n & I_n     \\
\end{bmatrix},
\end{equation}
and let $\varepsilon_l(t):=[\xi_l^T(t),\zeta_l^T(t)]^T=Q^{-1}e_l(t).$
Clearly, $\xi_l=x_l-x_{\sigma_i+1}$ and $\zeta_l=\eta_l-\eta_{\sigma_i+1}- x_l+x_{\sigma_i+1}$.
By \eqref{eq:CS_3}, one can obtain the following dynamic equations:
\[\begin{split}
\dot \xi_l(t) &= (A_i+B_i K_i) \xi_l(t)+ B_i K_i \zeta_l(t),\\
\dot \zeta_l(t)&= A_i\zeta_l(t) -c\left[c_i\sum_{k\in\cC_i}(b_{lk}-b_{\sigma_i+1,k}) \zeta_k(t)+\sum_{k\notin\cC_i}(b_{lk}-b_{\sigma_i+1,k})  \zeta_k(t)\right],
\end{split}
\]
for $l\in\cC_i$ and $l\neq \sigma_i+1$, $i=1,\ldots,N$. Since $K_i$ stabilizes $(A_i, B_i)$, the variable $\varepsilon_l(t)$ tends to zero as $t\rightarrow \infty$  if and only if $\zeta_l(t)$ tends to zero. Denote  $\zeta(t)=[\zeta_{\sigma_1+2}^T(t),\dots,\zeta_{\sigma_1+l_1}^T(t),
\cdots,\zeta_{\sigma_N+2}^T(t),\ldots,\zeta_{\sigma_N+l_N}^T(t)]^T$, which evolves with the following differential equation
\begin{equation} \label{eq:CS_10}
\dot \zeta(t)= \left(\hat \bA-c \hat \cL_c\otimes I_n\right)\zeta(t).
\end{equation}
Clearly, $\zeta(t)$ and every $\varepsilon_l(t)$  (hence every $e_l(t)$) all converge to zero if and only if $\hat \bA-c \hat \cL_c\otimes I_n$ is Hurwitz. That is, we have shown that $\lim_{t\rightarrow \infty}\|x_l(t)-x_k(t)\|=0$ and $\lim_{t\rightarrow \infty}\|\eta_l(t)-\eta_k(t)\|=0$, $\forall l,k\in\cC_i$, $i=1,\ldots,N$.

Next, we prove that $\eta_l(t)$ for any $l\in\cI$ vanishes as $t\rightarrow \infty$. To this end, for each $i=1,\ldots,N$, let $\eta_{i}(t)$ be the solution of $\dot{\eta}_{i}(t)= (A_i+B_iK_i)\eta_{i}(t)$ with an arbitrary initial value $\eta_{i}(0)$. Since $\sum_{k\in\cC_j}b_{lk}=0$ $\forall l\in\cI$ by Assumption \ref{assump:zero_row_sums}, we have that
\begin{align*}
\dot{\eta}_{i}(t)
&= (A_i+B_iK_i)\eta_{i}(t)\\
&= (A_i+B_iK_i)\eta_{i}(t) -c \left[c_i(\sum_{k\in\cC_i}b_{lk})(\eta_{\sigma_i+1}- x_{\sigma_i+1})\right.\\
&\left.\quad+\sum_{j=1,j\neq i}^{N}(\sum_{k\in\cC_j}b_{lk})(\eta_{\sigma_j+1}- x_{\sigma_j+1})\right],
\end{align*}
for any $l\in\cC_i$. Subtracting the above from \eqref{sys:controllaws_b} yields
\[\begin{split}
\dot{\eta}_{l}(t)-\dot{\eta}_{i}(t)
&= (A_i+B_iK_i)(\eta_{l}(t)-\eta_{i}(t))-c\left(c_i\sum_{k\in\cC_i}b_{lk}\zeta_k+ \sum_{j=1,j\neq i}^{N}\sum_{k\in\cC_j}b_{lk}\zeta_k\right).
\end{split}\]
The above system is exponentially stable and driven by inputs which all converge to zero exponentially fast. Therefore, for any $\eta_l(0), l\in\cI$, we have $\eta_l(t)\rightarrow \eta_{i}(t)\rightarrow 0$ $\forall l\in\cC_i$, as $t\rightarrow \infty$.

Lastly, we show that inter-cluster state separations can be achieved for any initial states $x_l(0)$'s by selecting $\eta_l(0)$'s properly. Given any set of $x_l(0)$, $l\in\cI$, choose $\eta_l(0)$, $l\in\cI$ such that $x_l(0)-\eta_l(0)=x_{\sigma_i+1}(0)-\eta_{\sigma_i+1}(0)$ for all $l\in\cC_i$, $i=1\ldots,N$, and $\lim \sup_{t\rightarrow \infty}\|e^{A_it}[x_l(0)-\eta_l(0)]-e^{A_jt}[x_l(0)-\eta_l(0)]\|\neq 0$ for any $i\neq j$. Considering the definition of $\zeta_l$ and the linear differential equation \eqref{eq:CS_10}, one has $x_l(t)-\eta_l(t)=x_{\sigma_i+1}(t)-\eta_{\sigma_i+1}(t)$ for all $t>0$. This together with \eqref{sys:controllaws_leaderless} lead to the following dynamics
$
\dot x_l(t)-\dot\eta_l(t)=A_i (x_l(t)-\eta_l(t)), \;\forall l\in\cC_i.
$
It follows that $x_l(t)=e^{A_it}[x_l(0)-\eta_l(0)]+\eta_l(t)\rightarrow e^{A_it}[x_l(0)-\eta_l(0)],\;\forall l\in\cC_i\;\; \textrm{as}\; t\rightarrow \infty.$
Therefore, $\lim \sup_{t\rightarrow\infty} \|x_{l}(t)-x_{k}(t)\|\neq 0$ $\forall l\in\cC_i$, $\forall k\in\cC_j$, $\forall i\neq j$.
This completes the proof.
\end{proof}

\subsection{Proof of Corollary \ref{thm:clustersyn_leaderless_hom}} \label{appendix:proof_clustersyn_leaderless_hom}
\begin{proof}
The proof for the necessity and sufficiency of \eqref{eq:CS_22b} is straightforward using the results in Lemma \ref{thm:lapalacian_reduce} and Theorem \ref{thm:clustersyn_leaderless}, and thus is omitted for simplicity. We only show that state separations are possible for any initial states $x_l(0)$, $l\in\cI$ by using the dynamic couplings even for systems with identical parameters.

Constellate the states $z_l(t)=[x_l^T(t),\eta_l^T(t)]^T$ of all $L$ agents to form $z(t):=[z_1^T(t),z_2^T(t),\ldots,z_{L}^T(t)]^T.$ It follows that
\begin{equation} \label{eq:linearmodel_leaderless_comp}
\dot z(t)=(I_L\otimes A_c-c\cL_c\otimes E) z(t),
\end{equation}
with
$A_{c}=\begin{bmatrix}
A & B K\\
0 & A+B K\\
\end{bmatrix}$ and $E=\begin{bmatrix}
 0 &0\\
-I_n &I_n
\end{bmatrix}.$  One can derive, after a series of manipulations, that
\begin{align*}
   z(t)\rightarrow \left[(\sum_{i=1}^N \mu_i \nu_i^T)\otimes e^{A_c t}\right]z(0),\;\;\textrm{as}\; t \rightarrow \infty,
\end{align*}
where each $\nu_i=[\nu_{i1},\ldots,\nu_{iL}]^T\in\RR^L$ is a left eigenvector of $\cL_c$ such that $\nu_i^T\cL_c=0$, $\nu_i^T\mu_i=1$, and $\nu_i^T\mu_j=0$, $\forall i\neq j$, with $\mu_1=[\ones_{l_1}^T,\zeros_{L-l_1}^T]^T,\mu_2=[\zeros_{l_1}^T,\ones_{l_2}^T,\zeros_{L-l_1-l_2}^T]^T,\ldots,\mu_N=[\zeros_{L-l_N}^T,\ones_{l_N}^T]^T$.
It then follows from the definitions of $z_l(t)$ and $z(t)$ that for all $l\in\cC_i,$
\begin{align*}
 x_l(t)& \rightarrow \sum_{k=1}^L \nu_{ik} [e^{At}x_k(0)+(e^{(A+BK)t}-e^{At})\eta_k(0)]\\
 &\rightarrow e^{At}\sum_{k=1}^L \nu_{ik} [x_k(0)-\eta_k(0)],\;\;\textrm{as}\; t \rightarrow \infty.
\end{align*}
Since $A$ is non-Hurwitz, $e^{At}$ is nonzero as $t \rightarrow \infty$. Then, for any set of initial states $x_l(0)$, $l\in\cI$, one can always find a set of $\eta_l(0)$, $l\in\cI$ such that
$\lim \sup_{t\rightarrow\infty} \|x_{l}(t)-x_{k}(t)\|\neq 0$ for any two agents $l\in\cC_i$ and $k\in\cC_j$, $i\neq j$.
This completes the proof.
\end{proof}

\subsection*{Acknowledgements} This work is supported by the Hong Kong RGC Earmarked Grant CUHK 14208314, Guangdong Natural Science Foundation (1614050001452), and Guangdong Science and Technology Program (2013B010406005, 2015B010128009).


\end{document}